\documentclass[12pt,cite,epsf,epsfig]{article}
\usepackage{epsfig}

\begin{document}

\author{S. Dev\thanks{dev5703@yahoo.com}, Shivani Gupta\thanks{shiroberts\_1980@yahoo.co.in}, Radha Raman Gautam\thanks{gautamrrg@gmail.com} and Lal Singh\thanks{lalsingh96@yahoo.com}}

\title{Near Maximal Atmospheric Mixing in Neutrino Mass Matrices with Two Vanishing Minors}
\date{\textit{Department of Physics, Himachal Pradesh University, Shimla 171005, India.}\\
\smallskip}

\maketitle
\begin{abstract}
In the flavor basis there are seven cases of two vanishing minors in the neutrino mass matrix which can accommodate the present neutrino oscillation data including the recent T2K data. It is found that two of these cases, namely $B_5$ and $B_6$ predict near maximal atmospheric neutrino mixing in the limit of large effective neutrino mass. This feature remains irrespective of the values of solar and reactor mixing angles. A non-zero reactor mixing angle is naturally accommodated in these textures. 

\end{abstract}

\section{Introduction}
During the past decade there has been considerable experimental development in the determination of neutrino masses and mixings \cite{1}. Recently, T2K experiment \cite{2} has given unambiguous hints of a relatively large 1-3 mixing angle. In this light, it is natural to look for models which, naturally, accommodate a non-zero value of reactor mixing angle while the atmospheric mixing angle remains near its maximal value. Recently many papers have appeared which reproduce the relatively large value of the reactor mixing angle \cite{3}.\\
There are mainly two approaches to explain neutrino mixings:
1) Mass independent textures \cite{4} which lead to mixing matrices independent of the eigenvalues. The most celebrated example of this category is the tribimaximal (TBM) \cite{5} scenario which has been derived from family symmetries and predicts a vanishing 1-3 mixing angle $\theta_{13} = 0$, maximal 2-3 mixing angle $\theta_{23} = \pi/4$ and 1-2 mixing angle $\theta_{12}$ = $\sin^{-1}(1/\sqrt{3})$. Non-zero $\theta_{13}$ can be accommodated in TBM and other similar models by considering deviations from symmetry.\\ 
2) Mass dependent textures which induce relations between mixing matrix elements and mass eigenvalues. Such textures naturally accommodate a non-zero $\theta_{13}$. Some examples of these are zero textures \cite{6}, vanishing minors \cite{7,8}, hybrid textures \cite{9}. Zero textures have been particularly successful in explaining both the quark and the lepton masses and mixings.\\ In this work we identify a class of mass dependent textures which supplemented with the assumption of a large value of effective neutrino mass $M_{ee}$ naturally predict near maximal $\theta_{23}$ and non-zero $\theta_{13}$. Recently, it was shown by Grimus \textit{et al} \cite{10} that near maximal atmospheric mixing is predicted for class B3 and B4 of two zero textures supplemented with the assumption of quasi degeneracy. We consider two vanishing minors of the neutrino mass matrix in the flavor basis together with the assumption of large $M_{ee}$. This assumption is well motivated by the extensive search for this parameter in the ongoing experiments. We found that the two cases of two vanishing minors viz. $B_5$ and $B_6$ [Table 1] in the limit of large $M_{ee}$ predict near maximal atmospheric mixing and this property holds irrespective of the values of solar and reactor mixing angles. The seesaw mechanism \cite{11} is regarded as the prime candidate for understanding the scale of neutrino masses. In the framework of type-I seesaw mechanism, the effective Majorana mass matrix $M_\nu$ is given by 
\begin{equation}
M_\nu = -M_D M_R^{-1} M_D^T
\end{equation}
where $M_D$ is the Dirac neutrino mass matrix and $M_R$ is the right-handed Majorana mass matrix. In the framework of type-I seesaw mechanism $M_\nu$ is a quantity derived from $M_D$ and $M_R$. Therefore, zeros of $M_D$ and $M_R$ have a deeper theoretical meaning. In a basis where $M_D$ is diagonal, the zeros of $M_R$ propagate as zero minors in $M_\nu$.
\begin{table}
\begin{small}
\begin{center}
\begin{tabular}{|c|c|}
\hline Class  & Zero Minors \\ 
\hline $A_1$ & $C_{3,3}$, $C_{3,2}$ \\ 
\hline $A_2$ & $C_{2,2}$, $C_{3,2}$ \\ 
\hline $B_3$ & $C_{3,3}$, $C_{3,1}$ \\ 
\hline $B_4$ & $C_{2,2}$, $C_{2,1}$ \\ 
\hline $B_5$ & $C_{3,3}$, $C_{1,2}$ \\ 
\hline $B_6$ & $C_{2,2}$, $C_{1,3}$ \\ 
\hline $D$ & $C_{3,3}$, $C_{2,2}$ \\ 
\hline
\end{tabular}
\caption{Experimentally allowed classes of two zero minors, here $C_{ij}$ denotes the zero minor corresponding to the $(ij)^{th}$ element of $M_\nu$.}
\end{center}
\end{small} 
\end{table}
\section{Symmetry realization}
In the basis where the charged lepton mass matrix is diagonal, there are fifteen possible two vanishing minors in $M_\nu$. Out of these fifteen only seven patterns  [Table-I] viz. $A_1$, $A_2$, $B_3$, $B_4$, $B_5$, $B_6$ and $D$ can accommodate the neutrino oscillation data. Of all the allowed two zero minors in the neutrino mass matrix only three cases $B_5$, $B_6$ and $D$ provide non-trivial zero minors, all other cases reduce to two zero textures when confronted with the neutrino oscillation data. We work in a basis where $M_D$ is diagonal ($M_D = diag(x,y,z)$), and the neutrino mixing arises solely from $M_R$. In this basis, a zero entry in $M_R$ propagates as a vanishing minor in the effective neutrino mass matrix $M_\nu$. Here, we focus on $B_5$ and $B_6$ class of vanishing minors. To obtain the classes $B_5$ and $B_6$ of neutrino mass matrices, we extend the Standard Model (SM) by adding three right-handed neutrino singlets $\nu_{R_i}$ and one scalar singlet $\chi$. In order to enable the seesaw mechanism for suppressing the neutrino masses $M_R$ must have the following structures for $B_5$ and $B_6$:
\begin{small}
\begin{equation}
M_R (B_5) = \left(
\begin{array}{ccc}
a& 0 &c \\
0 &d &e \\
c & e & 0
\end{array}
\right), \ M_R (B_6) = \left(
\begin{array}{ccc}
a & b & 0 \\
b &0 & e \\
0 & e & f
\end{array}
\right)
\end{equation}
\end{small}
leading to the following effective neutrino mass matrices through the seesaw mechanism
\begin{footnotesize} 
\begin{equation}
M_\nu (B_5) = \frac{1}{c^2 d+a e^2} \left(
\begin{array}{ccc}
e^2 x^2& -cexy &cdxz \\
-cexy &c^2 y^2 &aeyz \\
cdxz & aeyz & -adz^2
\end{array}
\right), \ M_\nu (B_6) =  \frac{1}{b^2 f+a e^2} \left(
\begin{array}{ccc}
e^2 x^2 & bfxy & -bexz \\
bfxy & -afy^2 & aeyz \\
-bexz & aeyz & b^2 z^2
\end{array}
\right).
\end{equation}
\end{footnotesize}
A general procedure for enforcing zero textures in arbitrary entries of the fermion mass matrices using abelian family symmetries has been outlined in \cite{12}. The symmetry realization of all the allowed one zero and two zero textures was recently presented in \cite{13}. For the symmetry realization of $B_5$ and $B_6$ textures of two zero minors we consider a small cyclic group $Z_3$ which corresponds to the minimal group since $Z_2$ leads to a non-diagonal charged lepton and Dirac neutrino mass matrix. Under $Z_3$ the SM Higgs doublet remains invariant and the leptonic fields are assumed to transform as:
\begin{eqnarray}
 D_{L_1}\rightarrow D_{L_1},& l_{R1}\rightarrow l_{R1},&  \nu_{R_1}\rightarrow \nu_{R_1},  \nonumber \\ D_{L_2}\rightarrow \omega D_{L_2},& l_{R2}\rightarrow \omega l_{R2}, & \nu_{R_2}\rightarrow  \omega \nu_{R_2},  \\ D_{L_3}\rightarrow \omega^2 D_{L_3},& l_{R3}\rightarrow  \omega^2 l_{R3}, & \nu_{R_3}\rightarrow \omega^2 \nu_{R_3}, \nonumber
\end{eqnarray}
where $\omega$ = $e^{i 2 \pi/3}$. Hence the bilinears $\overline{D}_{L_j} l_{R_k}$ and $\overline{D}_{L_j}\nu_{R_k}$, relevant for $M_l$ and $M_D$ transform as
\begin{small} 
\begin{equation}
\overline{D}_{L_j} l_{R_k} \sim \overline{D}_{L_j}\nu_{R_k} \sim \left(
\begin{array}{ccc}
1& \omega &\omega^2 \\
\omega^2 &1 &\omega \\
\omega& \omega^2&1
\end{array}
\right).
\end{equation}
\end{small}
The SM Higgs doublet remains invariant under $Z_3$ leading to diagonal $M_l$ and $M_D$. The bilinear $\nu_{R_j} \nu_{R_k}$ relevant for $M_R$ transforms as
\begin{small} 
\begin{equation}
\nu_{R_j} \nu_{R_k} \sim \left(
\begin{array}{ccc}
1& \omega &\omega^2 \\
\omega &\omega^2 &1 \\
\omega^2& 1& \omega
\end{array}
\right).
\end{equation}
\end{small}
We assume a scalar singlet $\chi$ transforming as $\chi \rightarrow \omega \chi$ for class $B_5$ and $\chi \rightarrow \omega^2 \chi$ for class $B_6$ which leads to the following $Z_3$ invariant Yukawa Lagrangians for classes $B_5$ and $B_6$:
\begin{small}
\begin{eqnarray}
-\mathcal{L}_{(B_5)} = Y_{11}^l \overline{D}_{L_1} \phi l_{R_1} +Y_{22}^l \overline{D}_{L_2} \phi l_{R_2} + Y_{33}^l \overline{D}_{L_3} \phi l_{R_3} + Y_{11}^{D} \overline{D}_{L_1} \tilde{\phi} \nu_{R_1} + Y_{22}^{D} \overline{D}_{L_2} \tilde{\phi} \nu_{R_2} + \ \ \ \ \nonumber \\ Y_{33}^{D} \overline{D}_{L_3} \tilde{\phi} \nu_{R_3} + \frac{Y_{13}^{M}}{2} \nu_{R_1}^T C^{-1} \nu_{R_3} \chi + \frac{Y_{22}^{M}}{2} \nu_{R_2}^T C^{-1} \nu_{R_2} \chi + \frac{M_{11}^{M}}{2} \nu_{R_1}^T C^{-1} \nu_{R_1} + \frac{M_{23}^{M}}{2} \nu_{R_2}^T C^{-1} \nu_{R_3} + H. c.
\end{eqnarray}
\begin{eqnarray}
-\mathcal{L}_{(B_6)} = Y_{11}^l \overline{D}_{L_1} \phi l_{R_1} +Y_{22}^l \overline{D}_{L_2} \phi l_{R_2} + Y_{33}^l \overline{D}_{L_3} \phi l_{R_3} + Y_{11}^{D} \overline{D}_{L_1} \tilde{\phi} \nu_{R_1} + Y_{22}^{D} \overline{D}_{L_2} \tilde{\phi} \nu_{R_2} + \ \ \ \ \nonumber \\ Y_{33}^{D} \overline{D}_{L_3} \tilde{\phi} \nu_{R_3} + \frac{Y_{12}^{M}}{2} \nu_{R_1}^T C^{-1} \nu_{R_2} \chi + \frac{Y_{33}^{M}}{2} \nu_{R_3}^T C^{-1} \nu_{R_3} \chi + \frac{M_{11}^{M}}{2} \nu_{R_1}^T C^{-1} \nu_{R_1} + \frac{M_{23}^{M}}{2} \nu_{R_2}^T C^{-1} \nu_{R_3} + H. c.
\end{eqnarray}
\end{small}
where $\tilde{\phi} = i \tau_2 \phi^*$. Next, we show how a large effective neutrino mass can arise in such a model. We note that $M_R$ contains two types of mass terms viz. 1) Bare mass term which does not need a scalar singlet and is invariant by itself. 2) Terms arising from Yukawa couplings to $\chi$. The scale of latter is restricted by the scale of $Z_3$ breaking while there is no such restriction on the bare mass term which can have a higher mass scale. It can be seen from eqn.(3) that the  $ee$ and $\mu \tau$ entries of $M_\nu$ have contributions to their numerators from $ee$ and $\mu \tau$ entries of $M_R$ which arise from the bare mass term. We assume the mass eigenvalues of $M_D$ to have same order of magnitude which leads to a large value of $ee$ and $\mu \tau$ entries of $M_\nu$ while the other elements of $M_\nu$ are suppressed, thus, leading to a large value of $M_{ee}$.
Since these textures are realized at the seesaw scale, the Renormalization Group (RG) evolution of the parameters of $M_\nu$ from the seesaw scale to the electroweak scale needs to be taken into account. It is well known that the RG effects are most prominent for the quasidegenerate mass spectrum which is precisely the case here due to the assumption of large $M_{ee}$. However, it is also known that zero minors in $M_\nu$, at a given energy scale, remain zero at any other energy scale at the one loop level \cite{7}. This is because the matrices at any two energy scales $\mu_1$ and $\mu_2$ are related by $M_\nu(\mu_1)=IM_{\nu}(\mu_2)I$, where $I$ is diagonal, positive and non singular. The operation of diagonal matrices from left and right on $M_\nu$ does not alter the zero minors of $M_\nu$ leading to zero minors in $M_\nu$ at any other scale. 
\section{Formalism}
We reconstruct the neutrino mass matrix in the flavor basis assuming neutrinos to be Majorana particles. In this basis, a complex symmetric neutrino mass matrix can be diagonalized by a unitary matrix $V$ as
\begin{equation}
M_{\nu}=VM_{\nu}^{diag}V^{T}
\end{equation}
where $M_{\nu}^{diag}$ = diag$(m_1,m_2,m_3)$. The  matrix $M_{\nu}$ can be parameterized in terms of three neutrino masses ($m_1, m_2, m_3$), three neutrino mixing angles ($\theta _{12}$, $\theta _{23}$, $\theta _{13}$) and the Dirac-type CP- violating phase $\delta$. The two additional phases $\alpha$ and $\beta$ appear if neutrinos are Majorana particles. The matrix 
\begin{equation}
V = UP
\end{equation}
where  \cite{14}
\begin{equation}
U= \left(
\begin{array}{ccc}
c_{12}c_{13} & s_{12}c_{13} & s_{13}e^{-i\delta} \\
-s_{12}c_{23}-c_{12}s_{23}s_{13}e^{i\delta} &
c_{12}c_{23}-s_{12}s_{23}s_{13}e^{i\delta} & s_{23}c_{13} \\
s_{12}s_{23}-c_{12}c_{23}s_{13}e^{i\delta} &
-c_{12}s_{23}-s_{12}c_{23}s_{13}e^{i\delta} & c_{23}c_{13}
\end{array}
\right)
\end{equation} with $s_{ij}=\sin\theta_{ij}$ and $c_{ij}=\cos\theta_{ij}$ and
\begin{small}
\begin{center}
$P = \left(
\begin{array}{ccc}
1 & 0 & 0 \\ 0 & e^{i\alpha} & 0 \\ 0 & 0 & e^{i(\beta+\delta)}
\end{array}
\right)$
\end{center}
\end{small}
is the diagonal phase matrix with
the two Majorana-type CP- violating phases $\alpha$, $\beta$ and Dirac-type CP-violating phase $\delta$. The matrix $V$ is called the neutrino mixing matrix or the Pontecorvo-Maki-Nakagawa-Sakata (PMNS) matrix  \cite{15}. Using Eq. (9) and Eq. (10), the neutrino mass matrix can be written as
\begin{equation}
M_{\nu}=U P M_{\nu}^{diag}P^{T}U^{T}.
\end{equation}
The CP violation in neutrino oscillation experiments can be described through a rephasing invariant quantity, $J_{CP}$ \cite{16} with $J_{CP}=Im(U_{e1}U_{\mu2}U_{e2}^*U_{\mu1}^*)$. In the above parametrization, $J_{CP}$ is given by
\begin{equation}
J_{CP} = s_{12}s_{23}s_{13}c_{12}c_{23}c_{13}^2 \sin \delta \   .
\end{equation}
The simultaneous existence of two vanishing minors in the neutrino mass matrix implies
\begin{eqnarray}
M_{\nu (pq)} M_{\nu (rs)}- M_{\nu
(tu)} M_{\nu (vw)}=0 \ , \\ M_{\nu (p'q')} M_{\nu (r's')}- M_{\nu
(t'u')} M_{\nu (v'w')}=0 \ .
\end{eqnarray}
These two conditions yield two complex equations viz.
\begin{eqnarray}
\sum_{l,k=1}^{3}(V_{pl}V_{ql}V_{rk}V_{sk}-V_{tl}V_{ul}V_{vk}V_{wk})m_{l}m_{k}=0 \ , \\
\sum_{l,k=1}^{3}(V_{p'l}V_{q'l}V_{r'k}V_{s'k}-V_{t'l}V_{u'l}V_{v'k}V_{w'k})m_{l}m_{k}=0 \ .
\end{eqnarray}
The above equations can be rewritten as
\begin{eqnarray}
m_1 m_2 A_3e^{2i\alpha} + m_2 m_3 A_1e^{2i(\alpha+\beta +\delta )}+m_3 m_1A_2e^{2i(\beta +\delta)}=0 \ , \\
m_1 m_2 B_3e^{2i\alpha} + m_2 m_3 B_1e^{2i(\alpha+\beta +\delta )}+m_3 m_1 B_2e^{2i(\beta +\delta)}=0 \ ,
\end{eqnarray}
where
\begin{eqnarray}
A_h=(U_{pl}U_{ql}U_{rk}U_{sk}-U_{tl}U_{ul}U_{vk}U_{wk})+(l\leftrightarrow k) \ ,\\ \nonumber
B_h=(U_{p'l}U_{q'l}U_{r'k}U_{s'k}-U_{t'l}U_{u'l}U_{v'k}U_{w'k})+(l\leftrightarrow k) \ ,
\end{eqnarray}
with $(h,l,k)$ as the cyclic permutation of (1,2,3). These two complex eqns.(18) and (19) involve nine physical parameters $m_{1}$, $m_{2}$, $m_{3}$, $\theta _{12}$, $\theta _{23}$, $\theta _{13}$ and three CP-violating phases $\alpha $, $\beta $ and $\delta $. The masses $m_{2}$ and $m_{3}$ can be calculated from the mass-squared
differences $\Delta m_{12}^{2}$ and $|\Delta m_{23}^{2}|$ using the relations
\begin{equation}
m_{2}=\sqrt{m_{1}^{2}+\Delta m_{12}^{2}} \ , \ \  m_{3}=\sqrt{m_{2}^{2}+|\Delta m_{23}^{2}|} \ 
\end{equation}
where $m_2 > m_3$ for Inverted Spectrum (IS) and $m_2 < m_3$ for Normal Spectrum (NS). 
Using the experimental inputs of the two mass-squared differences and the three mixing angles we can constrain the other parameters. Thus, in the two complex eqns.(18) and (19) we are left with four unknown parameters $m_1$, $\alpha$, $\beta$ and $\delta$ which are, obviously, correlated. Simultaneously solving eqns.(18) and (19) for the two mass ratios, we obtain
\begin{small}
\begin{equation}
\frac{m_1}{m_2}e^{-2i\alpha }=\frac{A_3 B_1 - A_1 B_3}{A_2 B_3 - A_3 B_2}
\end{equation}
\end{small}
and
\begin{small}
\begin{equation}
\frac{m_1}{m_3}e^{-2i\beta }=\frac{A_2 B_1 - A_1 B_2 }{A_3 B_2 - A_2 B_3}e^{2i\delta } \ .
\end{equation}
\end{small}
The mass ratios for class $B_5$ to first order in $s_{13}$ are given by
\begin{equation}
\frac{m_1}{m_2}e^{-2i\alpha } \approx 1+\frac{ s_{13} s_{23}
   \left(c_{23}^2 e^{- i \delta}+
   s_{23}^2 e^{i \delta} \right)}{c_{12} c_{23}^3 s_{12}}
\end{equation}
and
\begin{equation}
\frac{m_1}{m_3}e^{-2i\beta } \approx -\frac{s_{23}^2 e^{2 i \delta}}{c_{23}^2}-\frac{c_{12} s_{13} s_{23}^3 \left(c_{23}^2 e^{- i \delta}+
   s_{23}^2 e^{i \delta}\right) e^{2 i \delta}}{c_{23}^5 s_{12}} \ .
\end{equation}
The mass ratios for class $B_6$ to first order in $s_{13}$ are
\begin{equation}
\frac{m_1}{m_2}e^{-2i\alpha } \approx 1- \frac{ s_{13} c_{23}
   \left(c_{23}^2 e^{i \delta}+
   s_{23}^2 e^{-i \delta} \right)}{c_{12} s_{23}^3 s_{12}}
\end{equation}
and
\begin{equation}
\frac{m_1}{m_3}e^{-2i\beta } \approx -\frac{c_{23}^2 e^{2 i \delta}}{s_{23}^2}+\frac{c_{12} s_{13} c_{23}^3 \left(c_{23}^2 e^{i \delta}+
   s_{23}^2 e^{-i \delta}\right) e^{2 i \delta}}{s_{23}^5 s_{12}} \ .
\end{equation}
In the case of zero textures there exists a permutation symmetry between different patterns \cite{17}. Similarly, there exists a permutation symmetry between patterns $B_5$ and $B_6$ of two zero minors which corresponds to the permutation in the 2-3 rows and 2-3 columns of $M_\nu$. The corresponding permutation matrix is given by
\begin{small} 
\begin{equation}
P_{23} = \left(
\begin{array}{ccc}
1&0&0\\
0&0&1\\
0&1&0\\
\end{array}
\right).
\end{equation}
\end{small}
The right-handed Majorana mass matrix $M_R$ for class $B_6$ can be obtained from $M_R$ for class $B_5$ by the transformation
\begin{equation}
M_R^{B_6} = P_{23}M_R^{B_5}P_{23}^T
\end{equation}
which after the seesaw gives
\begin{equation}
M_{\nu}^{B_6} = P_{23}M_{\nu}^{B_5}P_{23}^T \ .
\end{equation}
This leads to the following relations between the parameters:
\begin{equation}
\theta_{12}^{B_6} = \theta_{12}^{B_5}, \ \theta_{13}^{B_6} = \theta_{13}^{B_5}, \ \theta_{23}^{B_6} = \frac{\pi}{2}-\theta_{23}^{B_5}, \ \delta^{B_6} = \delta^{B_5} - \pi \ .
\end{equation} 
The magnitude of the two mass ratios in eqns.(22, 23), is given by
\begin{equation}
\rho=\left|\frac{m_1}{m_3}e^{-2i\beta }\right| ,
\end{equation}
\begin{equation}
\sigma=\left|\frac{m_1}{m_2}e^{-2i\alpha }\right| .
\end{equation}
 while the CP- violating Majorana phases $\alpha$ and $\beta$ are given by
 \begin{small}
\begin{equation}
\alpha =-\frac{1}{2}arg\left(\frac{A_3 B_1 - A_1 B_3}{A_2 B_3 - A_3 B_2}\right),
\end{equation}
\end{small}
\begin{small}
\begin{equation}
\beta =-\frac{1}{2}arg\left(\frac{A_2 B_1 - A_1 B_2 }{A_3 B_2 - A_2 B_3}e^{2i\delta }\right).
\end{equation}
\end{small}
Since, $\Delta m_{12}^{2}$ and $|\Delta m_{23}^{2}|$ are known experimentally, the values of mass ratios $(\rho,\sigma)$ from eqns.(32) and (33) can be used to calculate $m_1$.
This can be done by inverting eqns.(21) to obtain the two values of $m_1$ viz.
\begin{small}
\begin{equation}
m_{1}=\sigma \sqrt{\frac{ \Delta
m_{12}^{2}}{1-\sigma ^{2}}} \ , \ \ 
m_{1}=\rho \sqrt{\frac{\Delta m_{12}^{2}+
|\Delta m_{23}^{2}|}{ 1-\rho^{2}}} \ .
\end{equation}
\end{small}
\section{Numerical Analysis}
The experimental constraints on neutrino parameters at 1, 2 and
3$\sigma$ \cite{18} are given below:
\begin{eqnarray}
\Delta m_{12}^{2}
&=&7.58_{(-0.26,-0.42,-0.59)}^{(+0.22,+0.41,+0.60)}\times
10^{-5}eV^{2}, \nonumber \\ |\Delta m_{23}^{2}| &=&
2.35_{(-0.09,-0.18,-0.29)}^{(+0.12,+0.22,+0.32)}\times 10^{-3}eV^{2},  \nonumber \\
\sin^2 \theta_{12}& =&0.312_{(-0.016,-0.032,-0.047)}^{(+0.017,+0.035,+0.052)},
\nonumber \\ \sin^2 \theta_{23}
&=&0.42_{(-0.03,-0.06,-0.08)}^{(+0.08,+0.18,+0.22)}, \nonumber \\
\sin^2 \theta_{13} &=&0.025_{(-0.007,-0.013,-0.020)}^{(+0.007,+0.016,+0.025)}.
\end{eqnarray}
The observation of  neutrinoless double beta (NDB) decay would signal lepton number violation and imply Majorana nature of neutrinos, for recent reviews see \cite{19,20}. The effective Majorana mass of the electron neutrino $M_{ee}$ which determines the rate of NDB decay is given by
\begin{equation}
M_{ee}= |m_1c_{12}^2c_{13}^2+ m_2s_{12}^2c_{13}^2 e^{2i\alpha}+ m_3s_{13}^2e^{2i\beta}|.
\end{equation}
Part of the Heidelberg-Moscow collaboration claimed a signal in NDB decay corresponding to $M_{ee} = (0.11 - 0.56) eV$ at 95$\%$ C. L. \cite{21}. This claim was subsequently criticized in \cite{22}. The results reported in \cite{21} will be checked in the currently running and forthcoming NDB experiments. 
There are large number of projects such as CUORICINO\cite{23}, 
CUORE \cite{24}, GERDA \cite{25}, MAJORANA \cite{26}, SuperNEMO \cite{27}, EXO \cite{28}, GENIUS \cite{29} which aim to achieve a sensitivity upto 0.01eV for $M_{ee}$. In the present work, we take the upper limit of $M_{ee}$ to be 0.5 $eV$ \cite{20}.
We vary the oscillation parameters within their known experimental ranges. However, the Dirac-type CP-violating phase $\delta$ is varied within its full range. The two values of $m_1$ obtained from the mass ratios $\rho$ and $\sigma$, respectively must be equal to within the errors of the oscillation parameters for the simultaneous existence of two vanishing vanishing minors in $M_\nu$.
The first step in the numerical analysis uses the information of the two known mass squared differences along with the constraint of two zero minors and large $M_{ee}$ to get predictions for the mixing angles. It is found that both the classes $B_5$ and $B_6$ predict a near maximal atmospheric mixing angle while the other two mixing angles remain unconstrained. The atmospheric mixing angle $\theta_{23}$ moves towards $\pi/4$ with increasing $M_{ee}$ as seen in Fig.(1) for class $B_5$ and Fig.(2) for class $B_6$. Thus classes $B_5$ and $B_6$ of two vanishing minors in $M_\nu$ naturally predict a near maximal atmospheric mixing angle. 
\begin{figure}
\begin{center}
{\epsfig{file=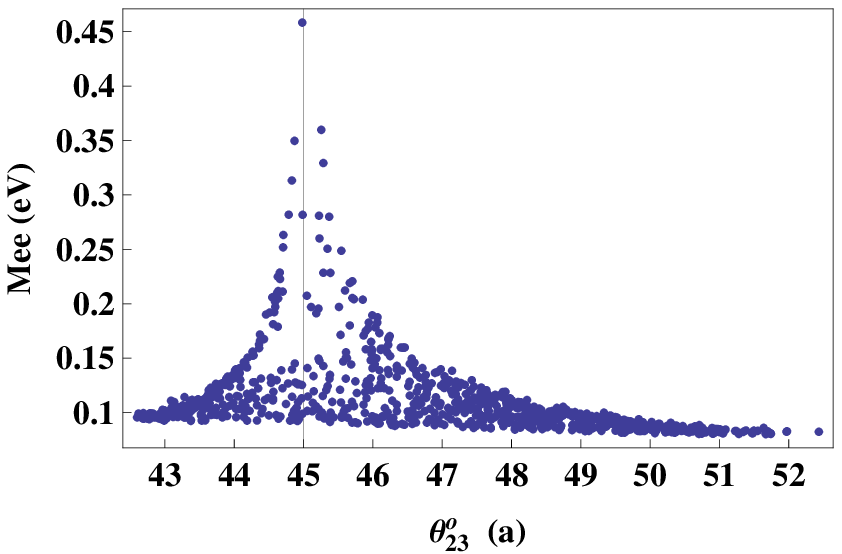, width=5.0cm, height=4.0cm} \  \epsfig{file=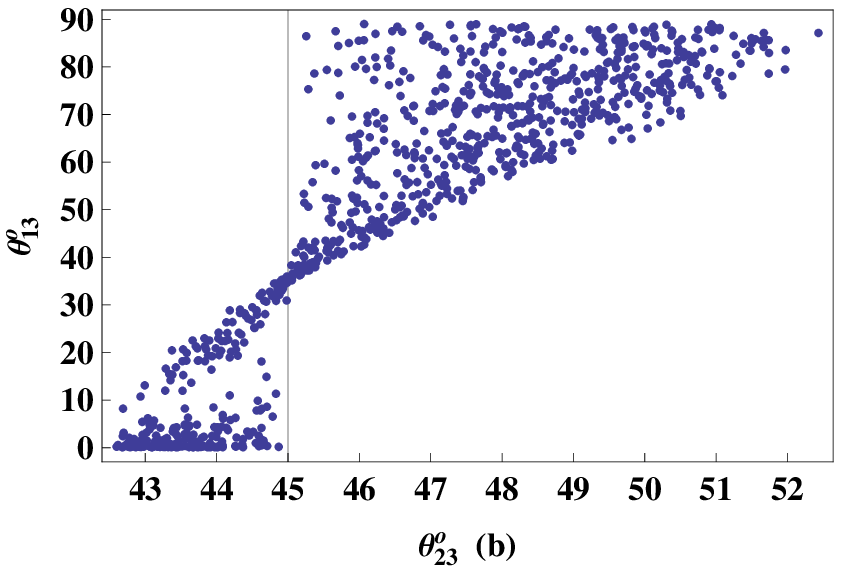, width=5.0cm, height=4.0cm} \ \epsfig{file=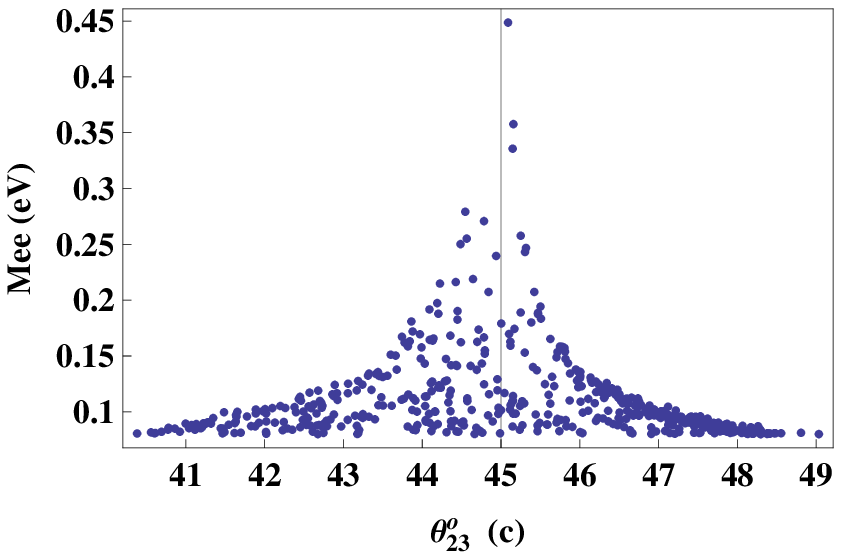, width=5.0cm, height=4.0cm} \ \epsfig{file=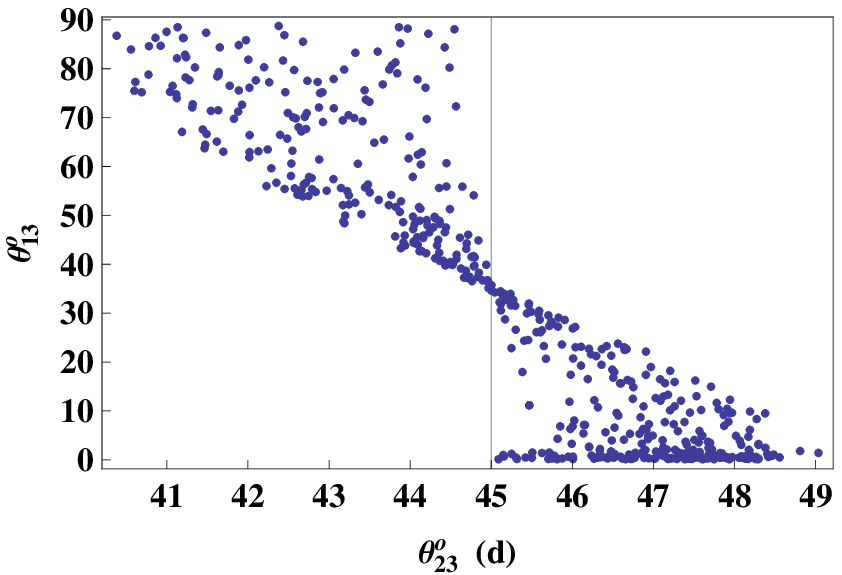, width=5.0cm, height=4.0cm}}
\end{center}
\caption{Correlation plots for class $B_5$, plots $(a)$, $(b)$ correspond to Normal Spectrum (NS) and plots $(c)$, $(d)$ correspond to Inverted Spectrum (IS).}
\end{figure}
\begin{figure}
\begin{center}
{\epsfig{file=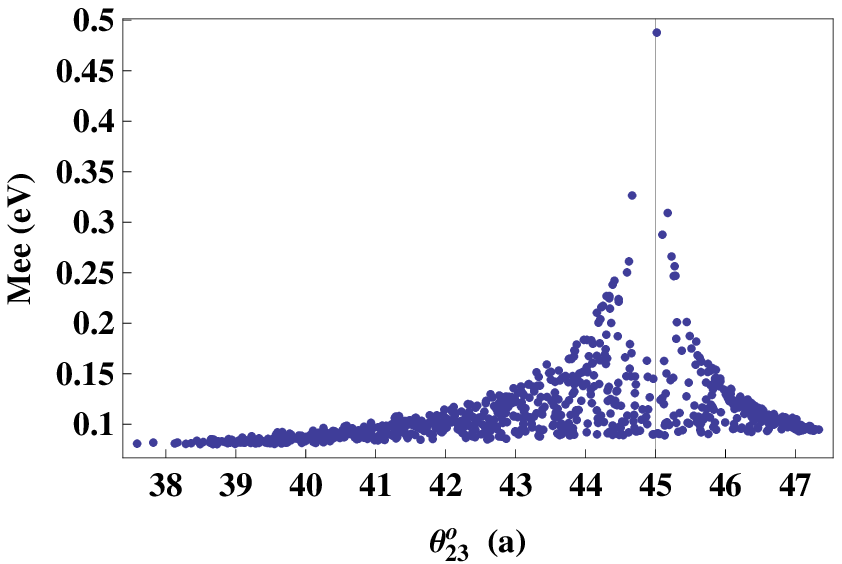, width=5.0cm, height=4.0cm} \  \epsfig{file=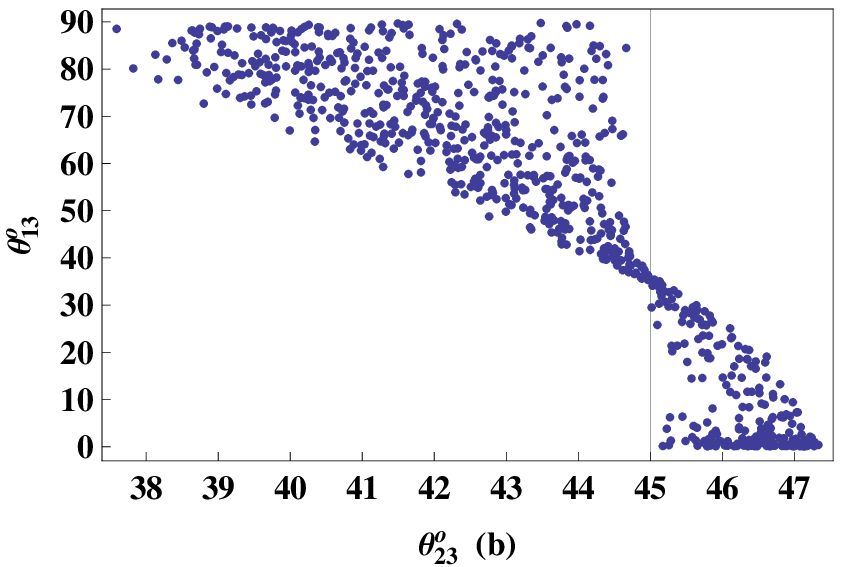, width=5.0cm, height=4.0cm} \ \epsfig{file=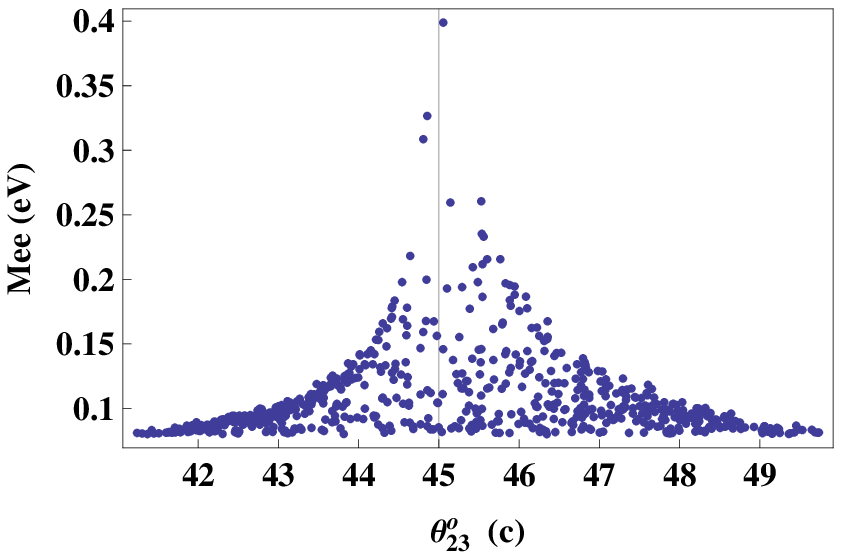, width=5.0cm, height=4.0cm} \ \epsfig{file=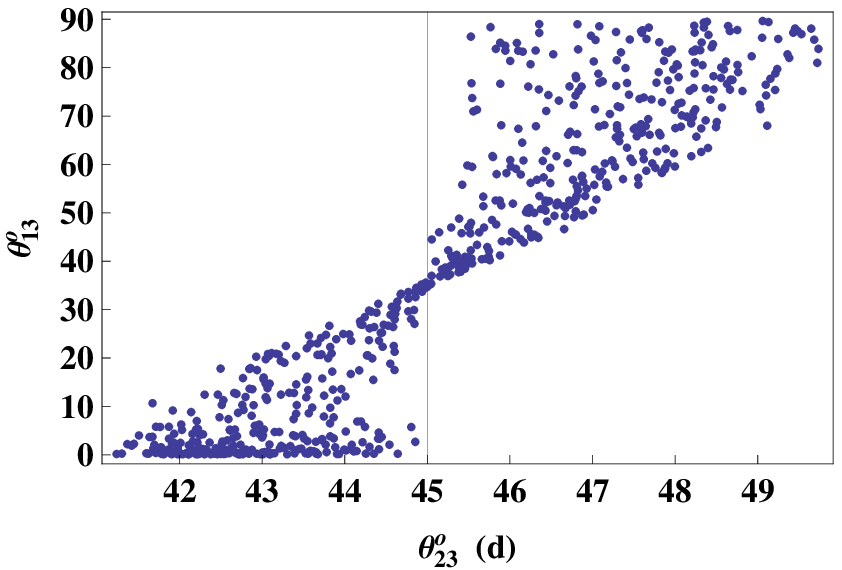, width=5.0cm, height=4.0cm}}
\end{center}
\caption{Correlation plots for class $B_6$, plots $(a)$, $(b)$ correspond to NS and plots $(c)$, $(d)$ correspond IS.}
\end{figure}
\begin{figure}
\begin{center}
{\epsfig{file=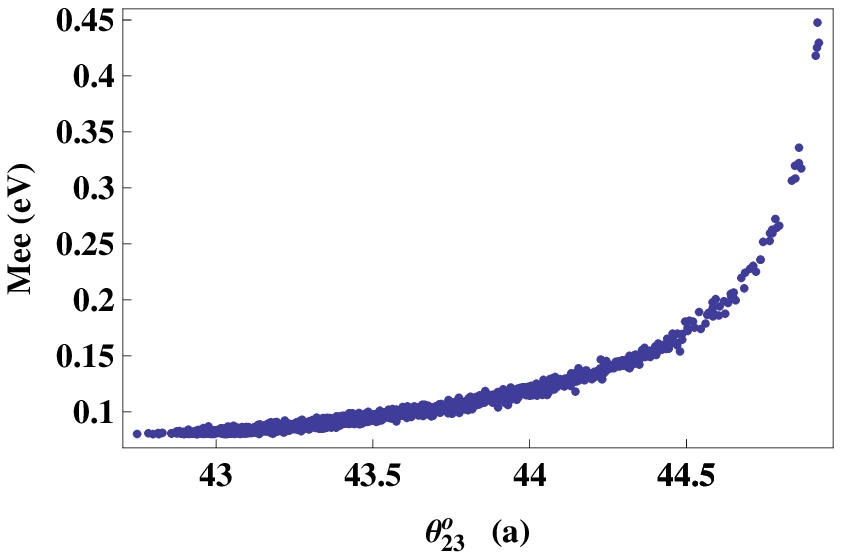, width=5.0cm, height=4.0cm} \  \epsfig{file=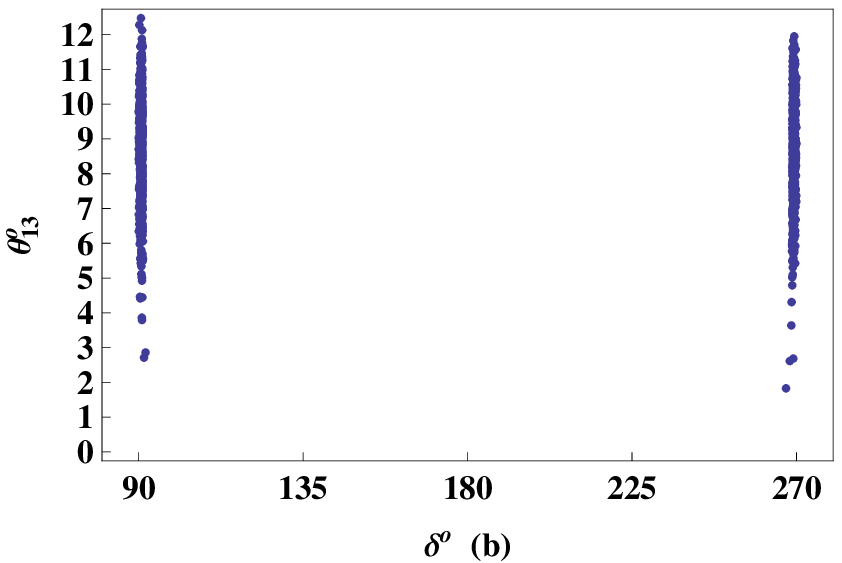, width=5.0cm, height=4.0cm} \ \epsfig{file=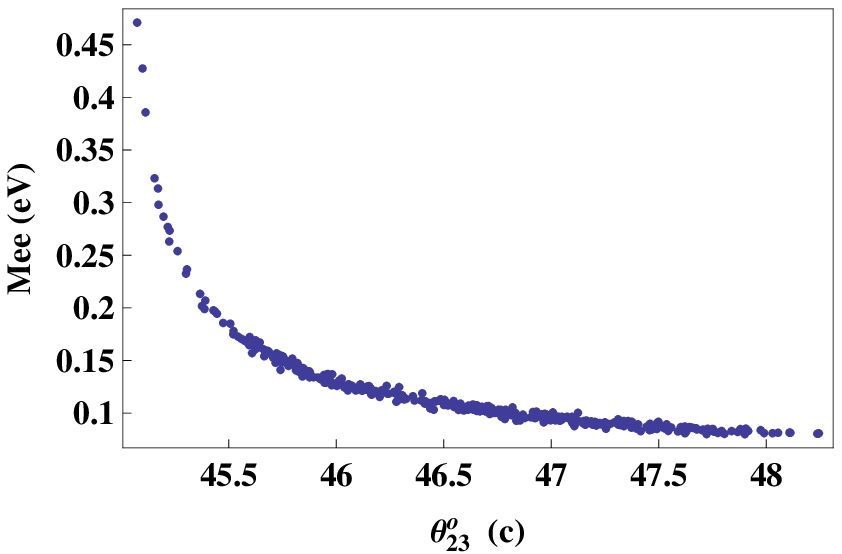, width=5.0cm, height=4.0cm} \ \epsfig{file=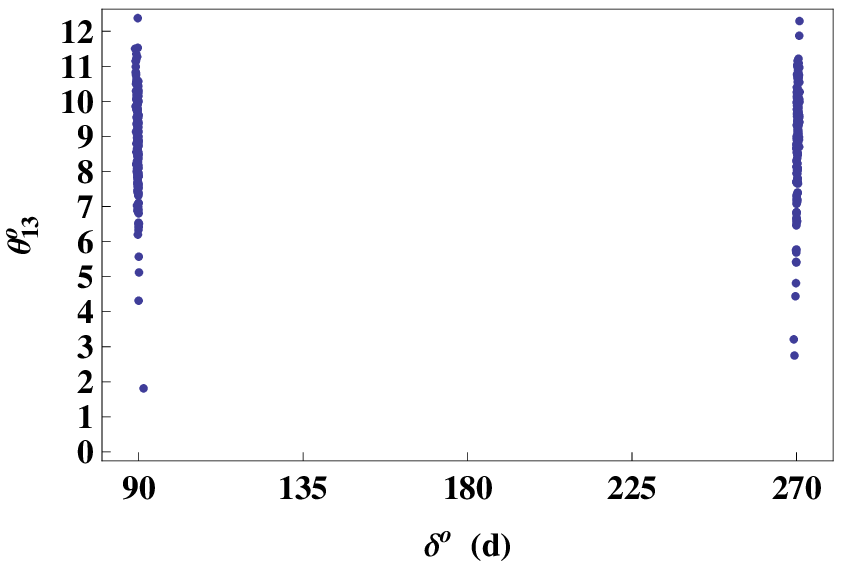, width=5.0cm, height=4.0cm}}
\end{center}
\caption{Correlation plots for class $B_5$, plots $(a)$, $(b)$ correspond to NS and plots $(c)$, $(d)$ correspond IS.}
\end{figure}
\begin{figure}
\begin{center}
{\epsfig{file=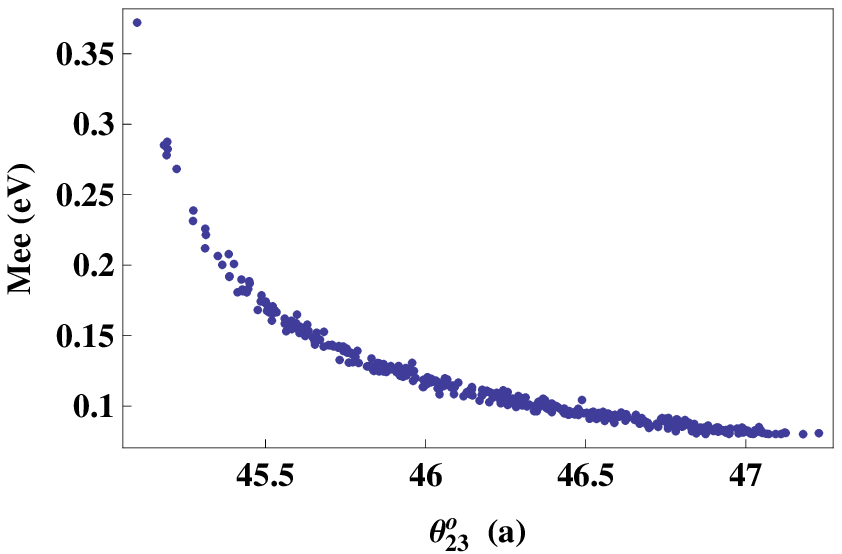, width=5.0cm, height=4.0cm} \  
\epsfig{file=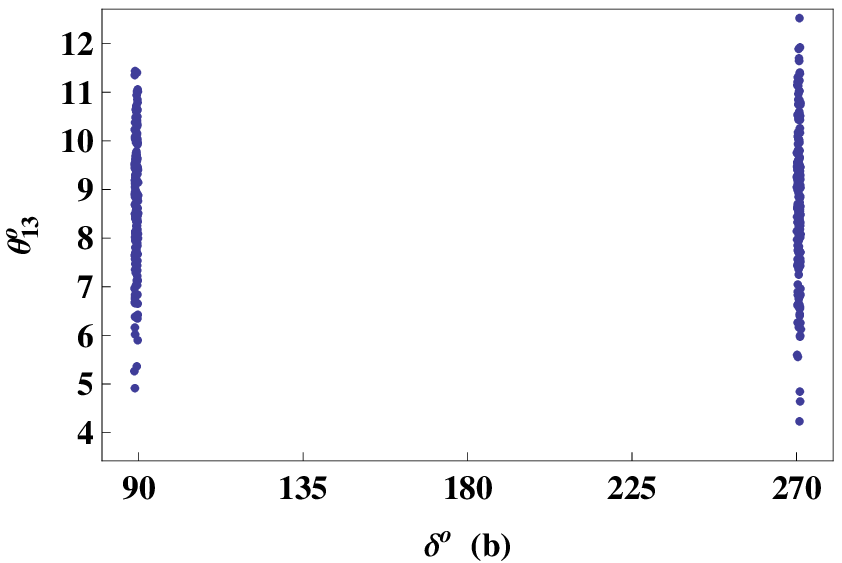, width=5.0cm, height=4.0cm} \ \epsfig{file=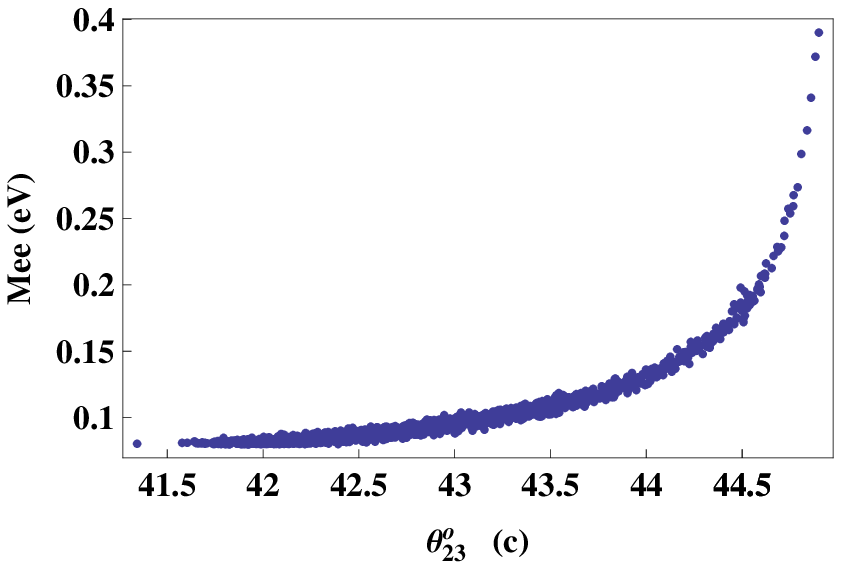, width=5.0cm, height=4.0cm} \
\epsfig{file=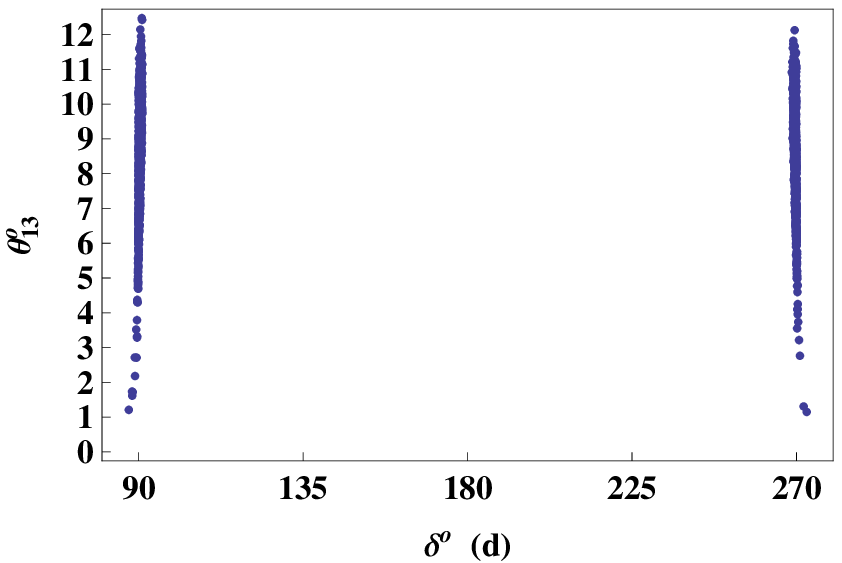, width=5.0cm, height=4.0cm}}
\end{center}
\caption{Correlation plots for class $B_6$, plots $(a)$, $(b)$ correspond to NS and Plots $(c)$, $(d)$ correspond IS.}
\end{figure}

The second step takes into account the experimental input on the three mixing angles including the recent T2K results on the reactor mixing angle. The results for $\theta_{23}$ are plotted in Fig.(3) for class $B_5$ and Fig.(4) for class $B_6$. Due to the relatively large value of $\theta _{13}$, the Dirac-type CP violating phase is almost fixed near $\pi/2$ or $3\pi/2$ predicting almost maximal CP violation for these textures. Fig.(5) and Fig.(6) show the correlation between the two Majorana-type phases and $M_{ee}$: the phases $\alpha$ and $\beta$ approach zero with increasing $M_{ee}$. As an example, we write the numerically estimated mass matrices for the pattern $B_5$, the matrices are obtained for the best fit values of $\Delta m_{12}^{2}$, $|\Delta m_{23}^{2}|$, $\theta_{12}$, $\theta_{13}$ given in eqn.(37). For NS we have $M_{ee} = 0.1129 eV$, $\theta_{23} = 43.902^o$, $\delta = 269.463^o$ and for IS we have $M_{ee} = 0.1272 eV$, $\theta_{23} = 46.058^o$, $\delta = 89.853^o$
\begin{footnotesize}
\begin{equation}
M_\nu^{B_5} (NS) = \left(
\begin{array}{ccc}
0.112956-0.000905 i &0.000205-0.002058 i&0.000018-0.000166 i\\
0.000205-0.002058 i&-0.000037-0.000008 i&-0.117485+0.002109 i\\
0.000018-0.000166 i&-0.117485+0.002109 i&-0.009482+0.000114 i\\
\end{array}
\right),
\end{equation}
\end{footnotesize}
\begin{footnotesize}
\begin{equation}
M_\nu^{B_5} (IS) = \left(
\begin{array}{ccc}
0.127210-0.001176 i &0.000147-0.002268 i&-0.000012+0.000177 i\\
0.000147-0.002268 i&-0.000040-0.000006 i&-0.122682+0.002478 i\\
-0.000012+0.000177 i&-0.122682+0.002478 i&0.009575-0.000154 i\\
\end{array}
\right).
\end{equation}
\end{footnotesize}
The numerical matrices for pattern $B_6$ can be obtained from above matrices with the operation of 2-3 permutation symmetry. The assumption of large $M_{ee}$ is testable in the ongoing and forthcoming experiments \cite{23,24,25,26,27,28,29} for NDB decay which will either confirm or rule out large $M_{ee}$ in the next few years. 
\begin{figure}
\begin{center}
{\epsfig{file=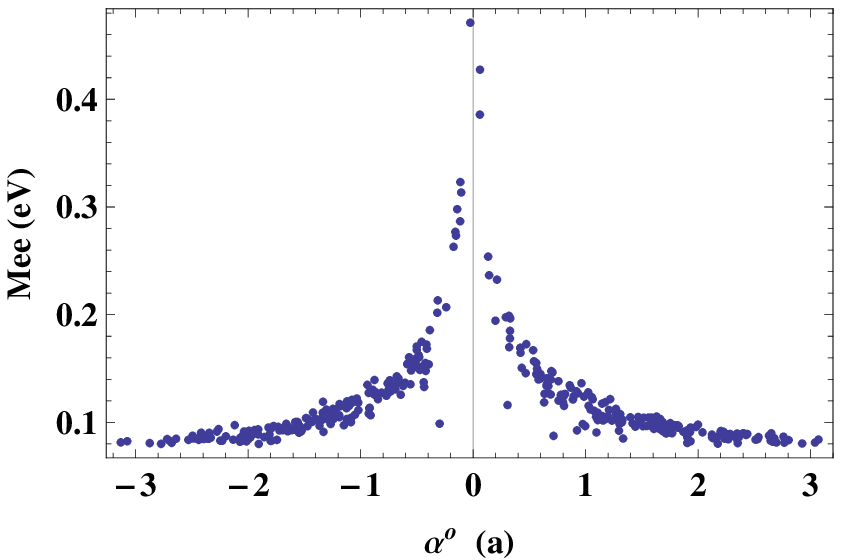, width=5.0cm, height=4.0cm} \  \epsfig{file=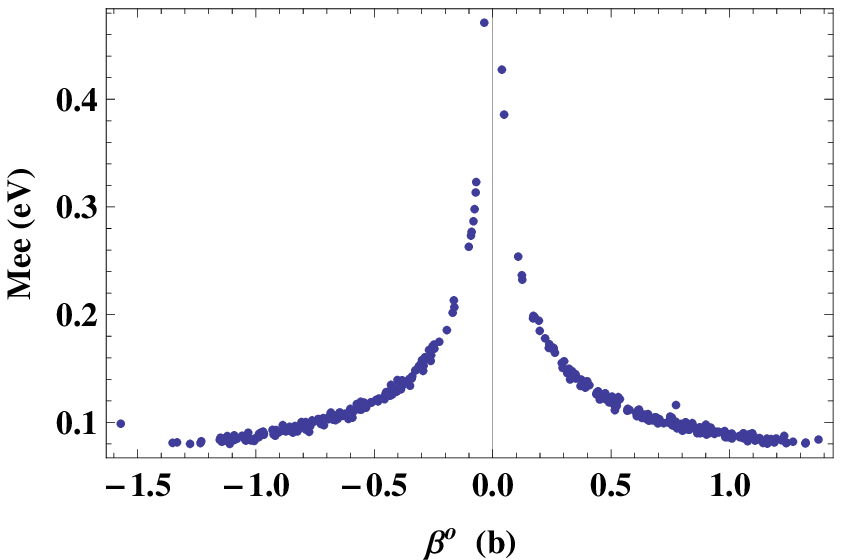, width=5.0cm, height=4.0cm}}
\end{center} 
\caption{Correlation plots of Majorana Phases with $M_{ee}$ for class $B_5$ (IS).}
\end{figure}
\begin{figure}
\begin{center}
{\epsfig{file=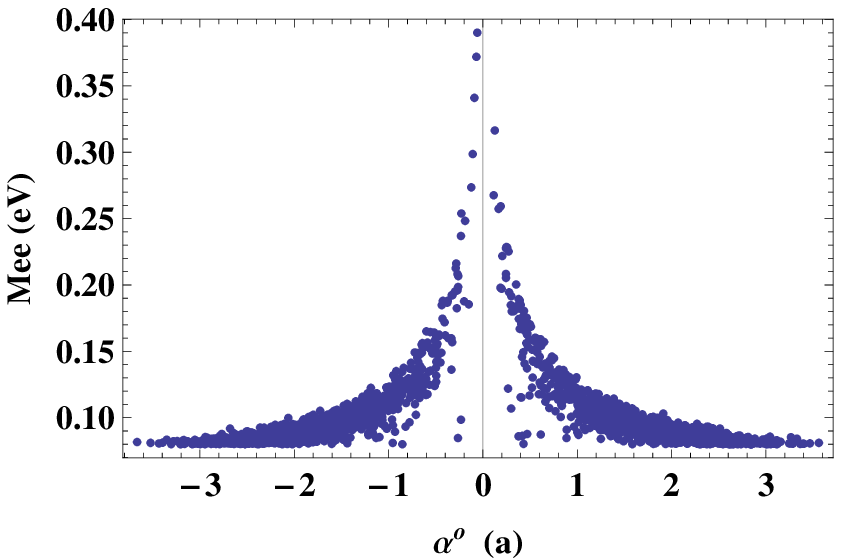, width=5.0cm, height=4.0cm} \  \epsfig{file=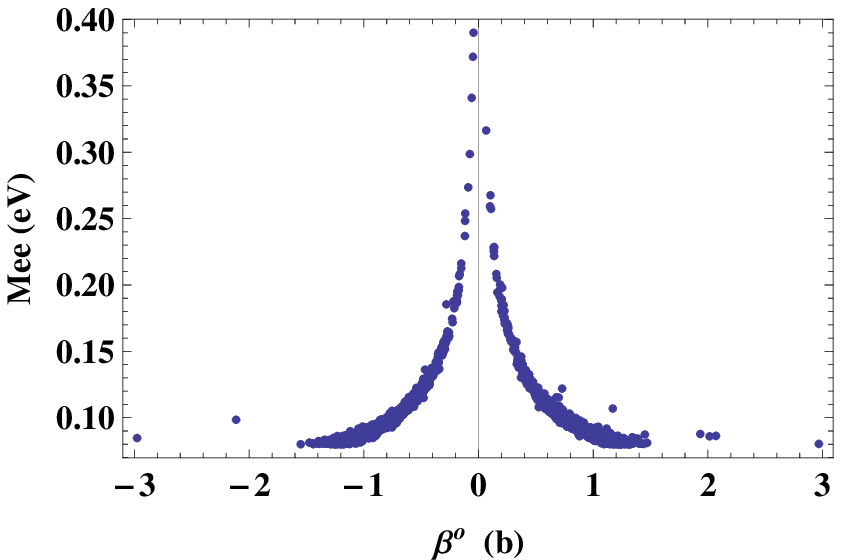, width=5.0cm, height=4.0cm}} 
\end{center}
\caption{Correlation plots of Majorana Phases with $M_{ee}$ for class $B_6$ (IS).}
\end{figure}
\section{Summary}
The recent results of the T2K experiment suggest a relatively large reactor mixing angle. Therefore, it is important to look for models naturally accommodating a non-zero value of reactor mixing angle while keeping the atmospheric mixing angle near maximal. In the present work, we studied the implications of class $B_5$ and $B_6$ of two zero minors in $M_\nu$ for large effective neutrino mass. In the context of type-I seesaw mechanism, taking $M_l$ and $M_D$ to be diagonal, the zeros of $M_R$ propagate as zero minors of $M_\nu$ and the origin of neutrino mixing is solely from $M_R$. We presented the symmetry realization of these patterns using a cyclic group $Z_3$. It was found that class $B_5$ and $B_6$ predict a near maximal atmospheric mixing angle in the limit of large $M_{ee}$. Furthermore, this prediction is independent of the values of the reactor and the solar mixing angles. The assumption of large $M_{ee}$ is testable in the ongoing experiments for NDB decay since the rate of NDB decay is proportional to $M_{ee}$. These experiments will either confirm or rule out a large value of $M_{ee}$ in the next few years. The atmospheric mixing angle approaches $\pi/4$ with the increasing value of $M_{ee}$. A reactor mixing angle equal to zero is not allowed in these textures, thus, naturally accommodating a non-zero $\theta_{13}$ as suggested by the recent results of the T2K experiment. Due to the relatively large value of $\theta_{13}$ the Dirac-type CP violating phase is fixed near $\pi/2$ or $3\pi/2$ predicting almost maximal CP violation for these textures.  
\\

\textbf{\textit{\Large{Acknowledgements}}}
The research work of S. D. and L. S. is supported by the University Grants
Commission, Government of India \textit{vide} Grant No. 34-32/2008
(SR). R. R. G. acknowledge the financial support provided by the Council for Scientific and Industrial Research (CSIR), Government of India.

\end{document}